\documentclass[conference]{IEEEtran}
\IEEEoverridecommandlockouts

\usepackage{cite}
\usepackage{amsmath,amssymb,amsfonts}
\usepackage{algorithmic}
\usepackage{graphicx}
\usepackage{textcomp}
\usepackage{inconsolata}
\usepackage{xcolor}
\usepackage{xspace}
\usepackage{verbatim}

\usepackage{fancyvrb}
\usepackage{todonotes}

\newcommand{\reasoner}{\xspace{\sf\small Solomon}\xspace}

\newcommand{\admin}{\xspace{\textit{System Administrator}}\xspace}
\newcommand{\lpaas}{\xspace{\textit{LPaaS}}\xspace}
\newcommand{\prolog}{\xspace{Prolog}\xspace}

\def\BibTeX{{\rm B\kern-.05em{\sc i\kern-.025em b}\kern-.08em
    T\kern-.1667em\lower.7ex\hbox{E}\kern-.125emX}}
\begin{document}
\vspace{-2mm}
\title{A Declarative Goal-oriented Framework\\for Smart Environments with LPaaS\thanks{Work partly supported by projects: \textit{GI\`O} funded by the Department of Computer Science of the University of Pisa, Italy; \textit{LiSCIo} (F4Fp-08-M30) funded by Fed4Fire+; \textit{CONTWARE} funded by the Conference of Italian University Rectors; and by the \textit{Orio Carlini Scholarship Programme 2020} funded by the GARR Consortium.}}

\author{\IEEEauthorblockN{Giuseppe Bisicchia, Stefano Forti, Antonio Brogi}
\IEEEauthorblockA{\textit{Department of Computer Science}
\textit{University of Pisa, Italy}
}
}

\maketitle

\begin{abstract}
Smart environments powered by the Internet of Things aim at improving our daily lives by automatically tuning ambient parameters (e.g. temperature, interior light) and by achieving energy savings through self-managing cyber-physical systems.
Commercial solutions, however, only permit setting simple target goals on those parameters and do not consider mediating conflicting goals among different users and/or system administrators, and feature limited compatibility across different IoT verticals.
In this article, we propose a declarative framework to represent smart environments, user-set goals and  customisable mediation policies to reconcile contrasting goals encompassing multiple IoT systems. An open-source Prolog prototype of the framework is showcased over two lifelike motivating examples. 
\end{abstract}


\section{Introduction}
\label{sec:intro}

\noindent
The Internet of Things (IoT) is continuously growing and becoming an integrated part of our daily lives with a plethora of new different applications (e.g. smart-environments, wearables, home appliances)~\cite{iotfuture,iotreview}, that show even capable of affecting our mood~\cite{iamhappy}. Among the new verticals the IoT is enabling, \textit{smart environments} are getting increasing attention from the market and the research community~\cite{smartenvreview,smarthomereview}. Indeed, they empower private and public ambients to self-manage cyber-physical systems (e.g. A/C, lights, plants watering) based on data from IoT sensors, triggering reactions enabled by IoT actuators. Besides their high potential to improve people's routines, these applications can also lead to a more sustainable energy and resource management~\cite{smartenvreview,smartenvreviewenergy}. 

Especially for those applications that include human goals in the self-management loop of smart environments, the problem of reconciling contrasting goals among different users emerges clearly~\cite{reasoningconflicts,rulebased}.
Colleagues sharing a room in a public building -- even for a limited amount of time -- can possibly express very different desiderata on the temperature and on the light intensity they prefer to experience while they work. To this end, many techniques have been proposed to reconcile such contrasting goals set by users or \textit{system administrators}, e.g. via fuzzy logic~\cite{fuzzyexpert}, multi-agent systems~\cite{masapproach,semioticmas} or neural networks~\cite{smartenvnn}.
However, most commercial solutions, such as IFTTT~\cite{ifttt} or Amazon Alexa~\cite{alexa}, only allow setting simple goals to be met by the IoT systems they manage and do not consider the possibility of mediating among contrasting objectives~\cite{surveyzambonelli}.

Additionally, despite being deployable out-of-the-box by their final users, existing commercial solutions show inherent limitations, mainly due to their proprietary nature. These limitations prevent them to be extended and from work \textit{across} IoT verticals enabled by different vendors. They also make it difficult to develop policies to mediate between users and administrator objectives, i.e. set local and global goals. Factually, two different types of conflict can arise:

\smallskip
\noindent\textit{User-user} -- Different users can set different goals on their desired state of the environment (e.g. on target temperature), 

\smallskip 
\noindent\textit{User-admin} -- The \admin can set global objectives that must be met (e.g. on maximum energy consumption, on law constraints), which may conflict with the user-set goals.
    
\smallskip
\noindent   
Even after reconciling the previous types of conflicts into one target state satisfying all set (user and/or global) goals, a final configuration of the actuators involved must also be determined. Indeed, given a final target state, we need to (\textit{a}) determine the correct configuration for each actuator acting on that state, and (\textit{b}) mediate between any conflicting configurations that a single actuator possibly receives.

\noindent
In this article, we propose a declarative methodology to specify customisable mediation policies for reconciling contrasting goals and actuator settings in smart environments. The methodology can solve contrasting goals by reasoning on the available IoT infrastructure and on (possibly contrasting) goals set by the users and by system administrators. 
The novel contribution mainly consists of:
\begin{enumerate}
    \item[(1)] a declarative framework to specify mediation policies for reconciling contrasting (user and/or global) goals and actuator settings in smart environments,
    \item[(2)] a Prolog prototype implementation of (1), \reasoner, provisioned as a REST service by relying on Logic Programming-as-a-Service (\lpaas)~\cite{lpaas}.
\end{enumerate}

\noindent
\reasoner tames the effects of the aforementioned types of conflict by allowing to flexibly specify \textit{ad-hoc} mediation policies for distinct zones of a smart-environment and possible conflicting settings of target actuators.  
Such policies can resolve conflicts (\textit{i}) among users' goals, (\textit{ii}) among users' and system administrator's goals, and \textit{(\textit{iii})} on actuators configuration. Last, but not least, the declarative nature of \reasoner makes it easy to write, maintain and extend arbitrary mediation policies encompassing multiple IoT verticals.

The rest of this article is organised as follows. After illustrating two motivating examples (Sect.~\ref{sec:examples}), we give some background on Prolog (Sect.~\ref{sec:background}). Then, we present our methodology for goal mediation and its prototype (Sect.~\ref{sec:methodology}), showcasing them over the first motivating example. The full prototype is subsequently assessed over the second motivating example (Sect.~\ref{sec:examples_retaken}). Finally, we discuss some closely related work (Sect.~\ref{sec:related}) before concluding (Sect.~\ref{sec:conclusions}).





\section{Motivating Examples}
\label{sec:examples}


\noindent
In this section, we illustrate two scenarios from smart environments to better highlight the need for reasoning solutions capable of mediating among contrasting goals and encompassing different IoT verticals. The first scenario considers a room in a \textit{Smart Home} (Sect. \ref{sec:smarthome}), while the second scenario, on a larger scale, considers a floor of a \textit{Smart Building} made of many offices and rooms (Sect.~\ref{sec:smartbuilding}). Both examples consider two main stakeholders:

\noindent\textit{User} --  a human or digital agent that can set goals on the ambient around them, aiming at creating the most comfortable environment for them to live in,

\noindent\textit{\admin} -- a human or digital agent that can define conflict resolution policies, and set global goals on the smart environment (e.g. on energy saving).


\subsection{\textit{Smart Home}}
\label{sec:smarthome}


\noindent
Consider a shared room in a student apartment, equipped with three lights -- a main light, a bed light and a corner light -- and an A/C system. In this case, depending on the time of day and the activity that is taking place (e.g. studying, watching a film, reading a book), different lighting configurations could be required. Conflicts might arise as, for instance, it can happen that Alice wants to watch a movie while Bob is still studying in the same room. Moreover, Alice might prefer to stay in a cool room (20$^{\circ}$C) while Bob prefers a warmer ambient (26$^{\circ}$C). 

Natural questions raised by the above scenario are:

\begin{itemize}
    \item[--]  \textit{Is it possible to find a configuration of the three lights which allows Alice and Bob to comfortably carry on their different activities?}
    \item[--]  \textit{Is it possible to find a configuration of the A/C system which mediates among the preferences of Bob and Alice on the environment temperature?}
\end{itemize}

\subsection{\textit{Smart Building}}
\label{sec:smartbuilding}


\noindent
Consider now the smart building floor sketched in Fig. \ref{fig:smartbuilding}, consisting of a West and East wings. The West wing is exposed to light most of the day while the East wing is less illuminated. 
In each wing, there are 5 rooms (4 single and one shared), the single rooms in pairs share the air conditioning system and the relative temperature sensor. Also, each room has a large light and a desk light and a brightness sensor. The shared rooms have two large lights and their air conditioning system as well as a temperature sensor and a brightness sensor. Additionally, the first single room in the East wing has also a small heater. Finally, each user has assigned a single room and has full access to both shared. Furthermore, targeting sustainability, company policies require that the temperature in the environment stays within 18$^{\circ}$C and 22$^{\circ}$C in autumn and winter, and between 24$^{\circ}$C and 28$^{\circ}$C in spring and summer. Also in these settings, some questions arise such as:

\begin{figure}[!ht]
    \centering
    \includegraphics[width=0.5\textwidth]{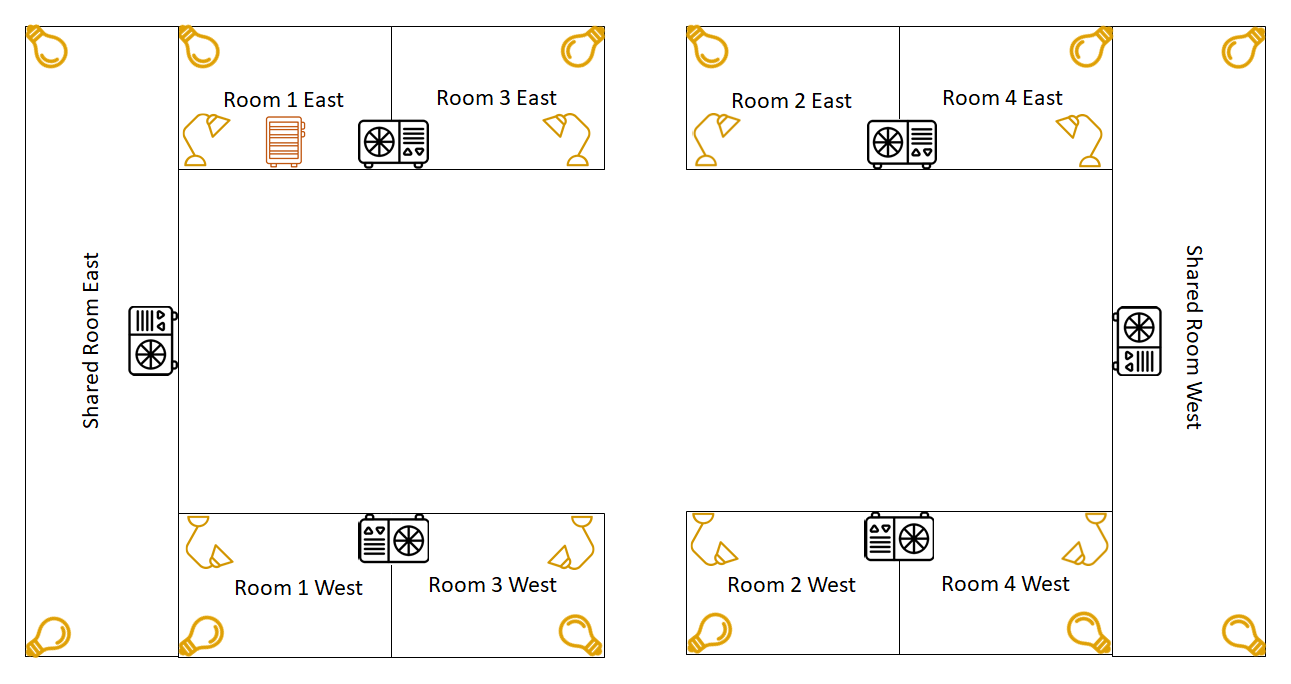}
    \caption{An example of \textit{Smart Building}}
    \label{fig:smartbuilding}
\end{figure}

\begin{itemize}
    \item[\textbf{Q1}]  \textit{How to describe the available Smart Building so that it is possible to apply \textit{ad-hoc} policies for the West and East wings?}
    \item[\textbf{Q2}] \textit{How can we specify policies to manage temperature and brightness in the different rooms of the building, handling conflicts so to ensure the comfort of its inhabitants and to meet sustainability policies?} 
    \item[\textbf{Q3}] \textit{Once a target state has been found for a specific room, how to determine suitable settings of the available (shared and non-shared) actuators to achieve it? }
\end{itemize}


\noindent All questions raised above highlight the need for novel models and methodologies to flexibly manage smart environments, such as the one we propose in this article. In the next section, we will detail our proposal by relying on the \textit{Smart Home} example. The \textit{Smart Building} example will be used instead in Sect.~\ref{sec:examples_retaken} to assess the methodology over a larger scale scenario.
\section{Background: Prolog \& LPaaS}
\label{sec:background}

\noindent
\prolog is a logic programming language as it is based on first-order logic. A Prolog program is a finite set of \textit{clauses} of the form:
%
\begin{Verbatim}[fontsize=\footnotesize, frame=single, framesep=1mm, framerule=0.1pt, rulecolor=\color{gray}]
a :- b1, ... , bn.
\end{Verbatim}
\vspace{-1.75mm}
%
\noindent 
stating that {\tt a} holds when {\tt b1} $\wedge\ \cdots\ \wedge$ {\tt bn} holds, 
where {\tt n}$\geq$0 and {\tt a}, {\tt b1}, ..., {\tt bn} are atomic literals.  Clauses can also contain inclusive disjunctions (i.e. logic ORs) among literals {\tt\small bi} and {\tt\small bj}, represented by {\tt\small bi; bj}.
Clauses with empty condition are also called \textit{facts}.
Prolog variables begin with upper-case letters, lists are denoted by square brackets, and negation by {\tt\small \textbackslash +}.
%

Recently, Calegari et al.~\cite{lpaas} have proposed to realise \lpaas, offering a flexible and lightweight inference engine as a REST service. 
%
\lpaas wraps a \prolog engine inside a REST server to manage incoming requests consistently. 
Such a service offers a well-defined API to upload Prolog facts and clauses that solve a domain-specific problem, to trigger reasoning over them, and to obtain computed solutions. LPaaS can be easily configured to handle stateful and stateless reasoning tasks, with static or dynamic knowledge bases.

Overall, \lpaas aims at enabling a plethora of different applications among ubiquitous and smart IoT systems, e.g. domestic robot assistants, smart kitchens to handle food supply based on user preferences, reasoning in sensor networks.
Particularly, \cite{lpaasiot} shows how \lpaas is well-suited for smart IoT applications and complex wireless networks, thanks to its high interoperability and customisation.
\section{Methodology and Prototype}
\label{sec:methodology}

\noindent
In this section, we illustrate \reasoner, a declarative framework featuring autonomic goal mediation in smart environments, in presence of multiple users.  
The framework is prototyped and open-sourced\footnote{Freely available at: {\tt\footnotesize https://github.com/di-unipi-socc/Solomon} } in Prolog, using \lpaas. We first give an overview of the architecture we foresee for \reasoner to be deployed (Sect.~\ref{sec:overview}), then we detail the model (Sect.~\ref{sec:model}) and methodology (Sect.~\ref{sec:reasoner}) underlying our framework.

\subsection{Overview}
\label{sec:overview}
    
\begin{figure}[!ht]
    \centering
    \includegraphics[width=0.45\textwidth]{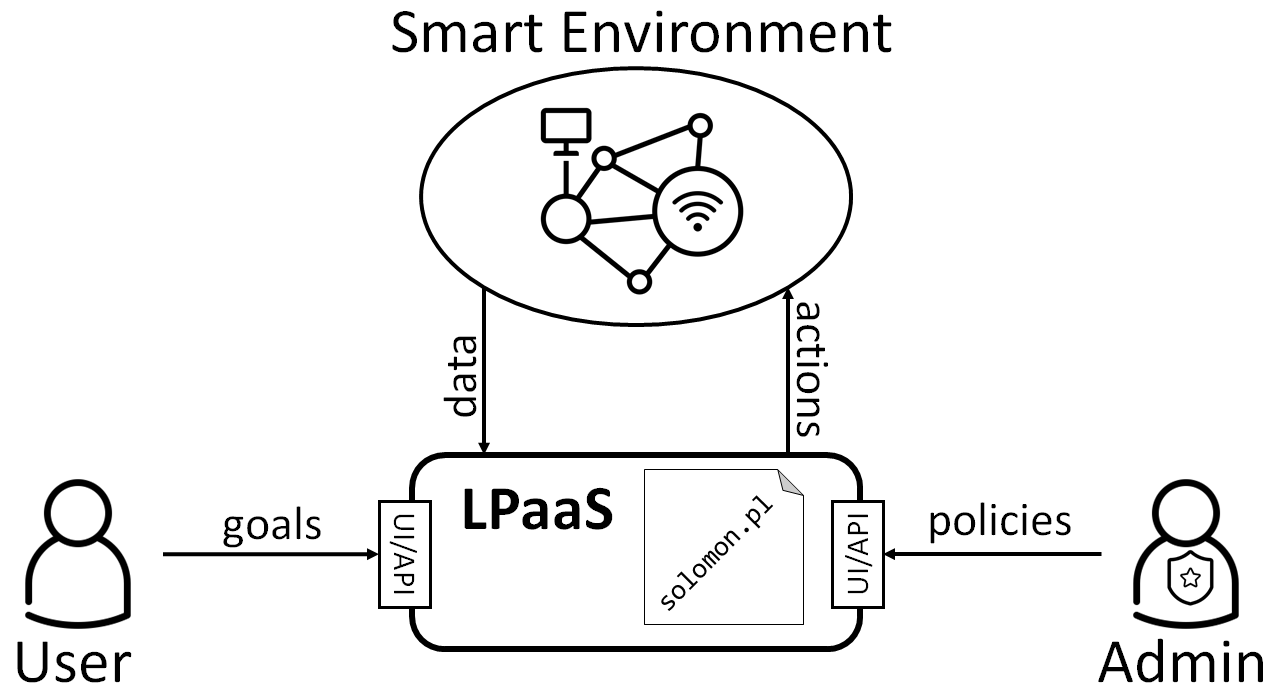}
    \caption{Blackbox view of \reasoner}
        \label{fig:boxology}
    \end{figure}

    \noindent
    Fig.~\ref{fig:boxology} gives a bird's-eye view of the architecture of \reasoner. 
    \reasoner interacts with a \textit{smart environment}, consisting of IoT sensors and actuators (or the services they are wrapped in). Indeed, \reasoner periodically receives updated \textit{data} from the sensors deployed in the smart environment, depending on which it can trigger suitable \textit{actions} for the available actuators.

\vspace{-3mm}
 
\vspace{-1.75mm}
\noindent

Both \textit{users} and \textit{system administrators} interact with \reasoner, through the \lpaas API or through available UIs. 
On one hand, users can declare \textit{goals} on the target state they wish to experience while being in the smart environment. On the other hand, system administrators can declare global mediation policies to solve user-user conflicts, to set global goals and solve user-admin conflicts, and to determine actuator configurations useful to reach a target state for the smart environment, after goal mediation.
Note that \reasoner is provisioned as a service, enabled by an \lpaas engine, which allows (\textit{i}) to easily integrate it with other pieces of software such as user interfaces (UIs) or mobile applications and (\textit{ii}) to deploy it either to Cloud or Edge servers, depending on the usage context.


    
    
    
    


\subsection{Model}
\label{sec:model}

\noindent\textit{Smart Environment. } To model smart environments, we first build up a dictionary of all types of environmental parameters we can monitor (via sensors) and/or act upon (via actuators). We call \textit{property types} the elements in such a dictionary, assuming they are declared as in
%
\begin{Verbatim}[fontsize=\footnotesize, frame=single, framesep=1mm, framerule=0.1pt, rulecolor=\color{gray}]
propertyType(TypeId).
\end{Verbatim}
\vspace{-1.75mm}
\noindent
where {\tt\small TypeId} is a literal value denoting the unique property type identifier. Given a {\tt\small propertyType} we can then define actuators and sensors that sense or operate on that. 

\noindent
Actuators are declared as in
\begin{Verbatim}[fontsize=\footnotesize, frame=single, framesep=1mm, framerule=0.1pt, rulecolor=\color{gray}]
actuator(ActuatorId, TypeId).
\end{Verbatim}
\vspace{-1.75mm}
\noindent
where {\tt\small ActuatorId} is the unique actuator identifier and {\tt\small TypeId} the associated property type. 

\noindent
Analogously, sensors are declared as in:
\begin{Verbatim}[fontsize=\footnotesize, frame=single, framesep=1mm, framerule=0.1pt, rulecolor=\color{gray}]
sensor(SensorId, TypeId).
\end{Verbatim}
\vspace{-1.75mm}
\noindent
where {\tt\small SensorId} is the unique sensor identifier and {\tt\small TypeId} is the associated property type. Environmental values monitored by each sensor are denoted by

\begin{Verbatim}[fontsize=\footnotesize, frame=single, framesep=1mm, framerule=0.1pt, rulecolor=\color{gray}]
sensorValue(SensorId, Value).
\end{Verbatim}
\vspace{-1.75mm}
\noindent
where {\tt\small SensorId} identifies the sensor and {\tt\small Value} is the last value it read.

\noindent \textbf{Example. } Based on the above, the shared room of the \textit{Smart Home} example of Sect. \ref{sec:smarthome} can be declared as in 

\begin{Verbatim}[fontsize=\footnotesize, frame=single, framesep=1mm, framerule=0.1pt, rulecolor=\color{gray}]
propertyType(light).
propertyType(temp).

sensor(brightness,  light).
sensor(temperature, temp).
sensorValue(brightness, 20).
sensorValue(temperature, 22).

actuator(smallLight, light).
actuator(mainLight, light).
actuator(cornerLight, light).
actuator(ac, temp).
\end{Verbatim}
\vspace{-1.75mm}
\noindent
where two sensors measure two different property types (i.e. temperature and brightness), having three lamps that can act on brightness and the AC system capable of changing the temperature. Please note that the current temperature settles at 22$^{\circ}$C and the brightness at $20$ out of $255$.
\hfill$\diamond$

\noindent
System administrators can divide smart environments into different \textit{zones}, which allow distinguishing which global policy to apply to specific sets of sensors and actuators: 

\begin{Verbatim}[fontsize=\footnotesize, frame=single, framesep=1mm, framerule=0.1pt, rulecolor=\color{gray}]
zone(ZoneId, MediationPolicy).
\end{Verbatim}
\vspace{-1.75mm}
\noindent
where {\tt\small ZoneId} is the unique zone identifier and {\tt\small MediationPolicy} is the unique identifier of the global management policy the zone is subject to. A zone groups one ore more \textit{property instances}, defining a set of actuators and a set of sensors that operate on a specific property type. 
 A property instance is declared as in

\begin{Verbatim}[fontsize=\footnotesize, frame=single, framesep=1mm, framerule=0.1pt, rulecolor=\color{gray}]
propertyInstance(ZId, PIId, TypeId, Actuators, Sensors).
\end{Verbatim}
\vspace{-1.75mm}
\noindent
where {\tt\small ZId} identifies the zone to which the instance belongs, {\tt\small PIId} is the property instance identifier, {\tt\small TypeId} is the {\tt\small propertyType} of {\tt\small PIId}, {\tt\small Actuators} is a list of actuators that operate on the property and {\tt\small Sensors} is a list of sensors that monitor it within the zone. All actuators and sensors in a given property instance must have the same property type. The identifier of a property instance is unique only within the zone, allowing for distinct zones to have instances with the same identifier.

 \noindent \textbf{Example. } 
The property instances of the \textit{Smart Home} example can be described by declaring a single {\tt\small livingroom} zone and, for instance, four property instances as in

\begin{Verbatim}[fontsize=\footnotesize, frame=single, framesep=1mm, framerule=0.1pt, rulecolor=\color{gray}]
zone(livingroom, _).
propertyInstance(livingroom, studyingLight, light, 
                 [cornerLight, mainLight], [brightness]).
propertyInstance(livingroom, movieLight, light,
                 [cornerLight, smallLight], [brightness]).
propertyInstance(livingroom, readingLight, light, 
                 [smallLight], [brightness]).
propertyInstance(livingroom, roomTemp, temp, 
                 [ac], [temperature]).
\end{Verbatim}
\vspace{-1.75mm}
\noindent It is worth noting that the first three property instances all refer to the {\tt\small light} property, grouping the brightness sensor with the lamps needed to realise different settings on such property, e.g. for studying ({\tt\small cornerLight} and {\tt\small mainLight}), watching a movie ({\tt\small cornerLight} and {\tt\small smallLight}), or reading a book ({\tt\small smallLight} only). The last property instance refers instead to the {\tt\small temp} property, grouping to the {\tt\small temperature} sensor and the A/C system (i.e. {\tt\small ac}).
\hfill$\diamond$

\noindent
\textit{Users and Goals. }
A \textit{user} is declared as in

\begin{Verbatim}[fontsize=\footnotesize, frame=single, framesep=1mm, framerule=0.1pt, rulecolor=\color{gray}]
user(UserId, AllowedZones).
\end{Verbatim}
\vspace{-1.75mm}
\noindent
where {\tt\small UserId} is the unique user identifier and {\tt\small AllowedZones} is the list of the zones on which the user can set goals. User \textit{goals} are declared as in

\begin{Verbatim}[fontsize=\footnotesize, frame=single, framesep=1mm, framerule=0.1pt, rulecolor=\color{gray}]
set(UId, ZId, PIId, Value).
\end{Verbatim}
\vspace{-1.75mm}
\noindent
where {\tt\small UId} is the user identifier, {\tt\small ZId} identifies a zone, {\tt\small PIId} is one of the property instances of the zone, and {\tt\small Value} is the goal expressed by the user on the property instance. 

 \noindent \textbf{Example. } Still following the \textit{Smart Home} scenario, Alice and Bob, and their goals on brightness and temperature are represented as per  

\begin{Verbatim}[fontsize=\footnotesize, frame=single, framesep=1mm, framerule=0.1pt, rulecolor=\color{gray}]
user(alice, [livingroom]).
set(alice, livingroom, movieLight, 20).
set(alice, livingroom, roomTemp, 20).
user(bob, [livingroom]).
set(bob, livingroom, studyingLight, 80).
set(bob, livingroom, roomTemp, 26).
\end{Verbatim}
\vspace{-1.75mm}
\noindent Alice aims at setting the {\tt\small movieLight} property instance to 20 out of 255 and the {\tt\small roomTemp} to 20$^{\circ}$C. Bob, on the contrary, wants to set the {\tt\small studyingLight} property instance to $80$ out of $255$, and the {\tt\small roomTemp} to 26$^{\circ}$C.
\hfill$\diamond$

\subsection{Reasoner}
\label{sec:reasoner}

\begin{figure*}[!ht]
    \centering
    \begin{Verbatim}[fontfamily=zi4, numbers=left, numbersep=5pt, fontsize=\footnotesize, numberblanklines=false, 
frame=single, 
framesep=1mm, framerule=0.1pt, rulecolor=\color{gray},  firstnumber=1,tabsize=1]
react(Requests, MediatedRequests, Actions) :- 
   getRequests(Requests, ValidRequests),  
   mediateRequests(ValidRequests, MediatedRequests), validMediation(MediatedRequests),
   associateActions(MediatedRequests, Actions), validActions(Actions).

getRequests(Requests, ValidRequests) :-
   findall((ZId, PIId, Value, UId), set(UId, ZId, PIId, Value), Requests),
   findall( (ZId, PIId, Value, UId), 
            ( member((ZId, PIId, Value, UId), Requests), user(UId, Zones), member(ZId ,Zones), validRequest(ZId, PIId, Value) ),
            ValidRequests).

validMediation(Reqs) :-
   sort(Reqs, OrderedReqs), 
   \+( ( member((Z,PI,V1), OrderedReqs), member((Z,PI,V2), OrderedReqs), dif(V1,V2) ) ),
   \+( ( member((Z,PI,V), OrderedReqs), \+( validRequest(Z,PI,V) ) ) ).

validActions(Actions) :-
   sort(Actions, OrderedActions),
   \+( ( member((A,V1), OrderedActions), member((A,V2), OrderedActions), dif(V1,V2) ) ),
   \+( ( member((A,V), OrderedActions), \+( validValue(A,V) ) ) ).
\end{Verbatim}
    \caption{\reasoner code.}
    \label{fig:reasoner}
\end{figure*}
\noindent
The model described up to now denotes the inputs that \reasoner receives from the smart environment it manages as well as from its users. 
Fig. \ref{fig:reasoner} lists the core code of \reasoner, which works in three main steps that constitute the top-down methodology of the proposed framework to determine a target state for a smart environment. Those steps are as follows:

\begin{enumerate}
\item it \textit{collects all user requests}\footnote{\texttt{findall(Template, Goal, Result)} finds all succesful solutions of \texttt{Goal} and collects the corresponding instantiations of \texttt{Template} in the list \texttt{Result}. If \texttt{Goal} has no solutions then \texttt{Result} is instantiated to the empty list.} that are currently submitted to the system ({\tt\small getRequests/2, line 2}) and extracts only those that are valid,
\item it \textit{mediates requests} referring to the same property instance by applying the mediation policies specified by the system administrator, so to determine a target state for each policy instance ({\tt\small mediateRequests/2, line 3}) by solving all user-user and user-admin conflicts,
\item it finally \textit{determines actions} (i.e. settings) for individual IoT actuators so to achieve the target state, by also resolving possible conflicting actions found for a single actuator ({\tt\small associateActions/2}, line 4).
\end{enumerate}

\noindent
Overall, the {\tt\small react/3} predicate (line 1) returns three lists: the list of all {\tt\small Requests}, the list of {\tt\small MediatedRequests} containing the target states for each property instance and the list of {\tt\small Actions} to perform to reach a final target state.
It is worth noting that, while the framework leaves complete flexibility to the system administrators in defining their own mediation policies, it also checks that inputs and outputs of each phase are well-formed (through predicates {\tt\small validMediation/1}, line 3, and {\tt\small validActions/1}, line 4). This guides the system administrators in their task of writing (formally) valid mediation policies.
%


\noindent\textit{collecting Requests.} First, \reasoner collects all the requests through {\tt\small getRequests/2} (line 2, lines 5--9), which determines two lists of tuples {\tt\small (ZId, PIId, Value, UId)}, where each tuple corresponds to a {\tt\small set(UId, ZId, PIId, Value)} with arguments rearranged for easier handling in later stages. The first list {\tt\small Requests} contains all current requests from users (line 6). The second one, {\tt\small ValidRequests}, only contains valid requests (line 7--9), i.e. by default\footnote{System administrators can easily extend the concept of \textit{valid request} by including further checks based on domain-specific knowledge, e.g. on the range of allowed values for a given property. This can be done by extending the {\tt\small validRequest/3} predicate exploited by {\tt\small getRequests/2} (line 8).}, requests for which the zone and the property instance exist, and the zone is among those the user associated with the request can set goals on. 

 \noindent \textbf{Example. } In the \textit{Smart Home} scenario,  querying
\begin{Verbatim}[fontsize=\footnotesize, frame=single, framesep=1mm, framerule=0.1pt, rulecolor=\color{gray}]
?- getRequests(Requests, ValidRequests).
\end{Verbatim}
\vspace{-1.75mm}
\noindent
returns the following:
\begin{Verbatim}[fontsize=\footnotesize, frame=single, framesep=1mm, framerule=0.1pt, rulecolor=\color{gray}]
Requests = ValidRequests, 
ValidRequests = [(livingroom, movieLight, 20, alice),  
                 (livingroom, studyingLight, 80, bob),  
                 (livingroom, roomTemp, 20, alice),  
                 (livingroom, roomTemp, 26, bob)].
\end{Verbatim}
\vspace{-1.75mm}
\noindent collecting all requests from Alice and Bob.
\hfill$\diamond$

\noindent\textit{Mediating Requests.} Valid requests are then passed to the {\tt\small mediateRequest/3} predicate (line 3) which can be flexibly and freely specified by the system administrator. 
The objective of this phase is to mediate between the possible conflicting goals of the users by determining one target value for each property instance. The {\tt\small mediateRequests/2} predicate outputs a list {\tt\small MediatedRequests} of such values for each property instance, in the form of triples {\tt\small (ZoneId, PropertyInstanceId, Value)}.
Then, the {\tt\small validMediation/1} predicate (lines 3, 10--13) checks that the list contains no duplicates (line 12) and that all requests are still valid after mediation (line 13). 


 \noindent \textbf{Example. }In our \textit{Smart Home} scenario, a possible {\tt\small mediateRequests/2} that simply averages user requests for a same property instance is as follows: 
\begin{Verbatim}[fontsize=\footnotesize, frame=single, framesep=1mm, framerule=0.1pt, rulecolor=\color{gray}]
mediateRequests(Requests, Mediated) :-
    groupPerPI(Requests, NewRequests),
    mediateRequest(NewRequests, Mediated).

mediateRequest([],[]).
mediateRequest([(Z,PI,Rs)|Reqs], [Mediated|OtherMedReqs]) :-
    mediatePI(Z,PI,Rs,Mediated),
    mediateRequest(Reqs, OtherMedReqs).

mediatePI(Z, PI, Ls, (Z, PI, Avg)) :-
    findall(V, member((V,_),Ls), Values), avg(Values,Avg).
\end{Verbatim}
\vspace{-1.75mm}
\noindent Input {\tt\small Requests} are first grouped per property instance by {\tt\small groupPerPI/2}, which returns a list of triples {\tt\small (Z,PI,Rs)} where {\tt\small Z} and {\tt\small PI} identify a property instance and {\tt\small Rs} is the list of requests that target it. 
By recursively scanning such list, {\tt\small mediateRequest/2} exploits {\tt\small mediatePI/4} to average 
all requests grouped for each property instance.

\noindent
By querying {\tt\small mediateRequests/4}, we obtain 

\begin{Verbatim}[fontsize=\footnotesize, frame=single, framesep=1mm, framerule=0.1pt, rulecolor=\color{gray}]
Mediated =  [(livingroom, movieLight, 20),  
             (livingroom, roomTemp, 23),
             (livingroom, studyingLight, 80)].
\end{Verbatim}
\vspace{-1.75mm}
\noindent which represents a target state where {\tt\small movieLight} and {\tt\small studyingLight} are set to 20 and 80 respectively, and {\tt\small roomTemp} to 23$^\circ$C, i.e. the average of Bob and Alice's goals.
\hfill$\diamond$

\noindent\textit{Determining Actions.} After obtaining a target state for each property instance, \reasoner generates a list of actions for available actuators to reach such target. 
An action is a pair {\tt\small (AId, Value)} where {\tt\small AId} is the identifier of an actuator and {\tt\small Value} is the value it need to be set to. The {\tt\small associatedActions/2} predicate (line 4) inputs a list of mediated requests and returns a list of actions, according to \admin policies. 

The {\tt\small validActions/1} predicate (lines 4, 14--17) checks whether there are no duplicate settings (line 18) and that obtained values are valid according to \admin policies (line 17), which can check if the configuration for an actuator can factually be implemented, using  {\tt\small validValue/1}.


 \noindent \textbf{Example. }A simple policy that computes the setting for each actuator by dividing the target value of a  {\tt\small propertyInstance} by the number of its actuators, is specified as in 
\begin{Verbatim}[fontsize=\footnotesize, frame=single, framesep=1mm, framerule=0.1pt, rulecolor=\color{gray}]
associateActions(Requests, ExecutableActions) :-
    actionsFor(Requests, Actions),            
    setActuators(Actions, ExecutableActions). 

actionsFor([],[]).
actionsFor([(Z, PI, V)|Reqs], Actions) :-
    propertyInstance(Z, PI, _, Actuators, _),              
    selectActionsForPI(Z, PI, V, Actuators, _, Actions1), 
    actionsFor(Reqs, Actions2), 
    append(Actions1, Actions2, Actions).   

selectActionsForPI(_, _, V, Actuators, _, Actions) :-
    length(Actuators, L),triggerAll(V, L, Actuators, Actions).

triggerAll(_, _, [], []).
triggerAll(V, L, [A|Actuators], [(A,VNew)|Actions]) :-
    VNew is V/L, triggerAll(V, L, Actuators, Actions).

setActuators(Actions, ExecutableActions) :- 
    setActuatorsWithMax(Actions, 0, 100, ExecutableActions).
\end{Verbatim}
\vspace{-1.75mm}
\noindent First, for each input {\tt\small requests}, a triple {\tt\small (Zone, PropertyInstance, TargetValue)}, {\tt\small actionsFor/2} gets the list of actuators of that specific {\tt\small propertyInstance}. Then, it  {\tt\small selectActionsForPI/6} computes the list of actions to be performed by dividing the target value for each {\tt\small propertyInstance} by the number of its actuators. Note that when an actuator belongs to more than one {\tt\small propertyInstance}s, {\tt\small setActuators/2} selects the highest value available cutting that value with a lower bound of 0 and an upper bound of 100. 
%

By querying {\tt\small associateActions/2} in the \textit{Smart Home} scenario, given the target state of the previous example, we obtain:
\begin{Verbatim}[fontsize=\footnotesize, frame=single, framesep=1mm, framerule=0.1pt, rulecolor=\color{gray}]
?- associateActions(Mediated, Actions).
Actions = [(ac, 23), (cornerLight, 40),  
           (mainLight, 40), (smallLight, 10)] 
\end{Verbatim}
\vspace{-1.75mm}
Note that the {\tt\small ac} actuator is set to 23$^\circ$C, i.e. the value of the target state. As for {\tt\small movieLight} and {\tt\small studyingLight}, being composed of several actuators, a further mediation happens. The target value of 20 for {\tt\small movieLight} is split across {\tt\small cornerLight} and {\tt\small smallLight}, setting each to 10. Analogously, the target value of 80 for the {\tt\small studyingLight} is split across {\tt\small mainLight} and {\tt\small cornerLight}, setting each to 40. The conflict on {\tt\small cornerLight}, being in both property instances, is solved by picking the maximum between 10 and 40, viz. 40.
\hfill$\diamond$

\section{Smart Building Example Retaken}
\label{sec:examples_retaken}

\noindent
In this section, we exploit \reasoner to answer the questions raised about the \textit{Smart Building} scenario of Sect.~\ref{sec:smartbuilding}.


\begin{figure*}[!ht]
    \centering
    \begin{Verbatim}[fontfamily=zi4, numbers=left, numbersep=5pt, fontsize=\footnotesize, numberblanklines=false, 
frame=single, 
framesep=1mm, framerule=0.1pt, rulecolor=\color{gray},  firstnumber=1,tabsize=2]
findValue(_, temp, _, TempValue, Value) :-
    season(S),
    (((S = winter ; S = autumn), (TempValue > 22, Value is 22; TempValue < 18, Value is 18; Value is TempValue));
    ((S = summer ; S = spring), (TempValue > 28, Value is 28; TempValue < 24, Value is 24; Value is TempValue))).

findValue(east, light, _, LightValue, Value) :-
    (LightValue > 255, Value is 255; LightValue < 100, Value is 100; Value is LightValue).

findValue(west, light, Brightness, LightValue, Value) :-
    ((Brightness > 100, (LightValue > 255, Value is 255; LightValue < 100, Value is 100; Value is LightValue));
    (LightValue > 255, Value is 255; LightValue < 180, Value is 180; Value is LightValue)).
    \end{Verbatim}
    \caption{{\tt\small findValue/4} implements global policies in the \textit{Smart Building}.}
    \label{fig:findValue}
\end{figure*}

The answer to \textbf{Q1}  is obtained by first specifying different {\tt\small zone(ZoneId, MediationPolicy)} facts for the rooms in the smart building (Fig. \ref{fig:smartbuilding}), as in

\begin{Verbatim}[fontsize=\footnotesize, frame=single, framesep=1mm, framerule=0.1pt, rulecolor=\color{gray}]
zone(room_E_1, east).
zone(room_E_2, east).
zone(room_E_3, east).
zone(room_E_4, east).
zone(room_W_1, west).
zone(room_W_2, west).
zone(room_W_3, west).
zone(room_W_4, west).
zone(commonRoom_E, east).
zone(commonRoom_W, west).
\end{Verbatim}
\vspace{-1.75mm}
\noindent
The {\tt\small east} and {\tt\small west} literals identify two different mediation policies, specified by \admin, to be applied to the property instances grouped under the zone. Such grouping can be obtained by specifying suitable {\tt\small propertyInstance(ZId, PIId, TypeId, Actuators, Sensors)} facts as, for instance, in 

\begin{Verbatim}[fontsize=\footnotesize, frame=single, framesep=1mm, framerule=0.1pt, rulecolor=\color{gray}]
propertyInstance(room_E_1, roomTemp, temp, 
                [acOdd_E, heater], [tempOdd_E]).
propertyInstance(room_E_1, roomLight, light, 
                [biglightRoom_E_1, smalllightRoom_E_1], 
                [lightRoom_E_1]).
propertyInstance(room_E_3, roomTemp, temp, 
                [acOdd_E], [tempOdd_E]).
propertyInstance(room_E_3, roomLight, light, 
                [biglightRoom_E_3, smalllightRoom_E_3], 
                [lightRoom_E_3]).
\end{Verbatim}
\vspace{-1.75mm}
\noindent that describes the sensors and actuators available in the \textit{Room~1} and \textit{Room~3} of the East wing. Note that the two rooms share the {\tt\small acOdd\_E} actuator for the A/C system and that \textit{Room~1} contains the {\tt\small heater} actuator that is not available in \textit{Room~3}.


Based on the knowledge representation above, we can now answer \textbf{Q2} by suitable implementations of {\tt\small mediateRequests/2}. Indeed, the \admin can easily declare mediation policies to solve user-user and user-admin conflicts in a context-aware manner. Such behaviour can be obtained through predicate {\tt\small mediatePI/4} (which is used by {\tt\small mediateRequests/2} as illustrated in Sect. \ref{sec:methodology}) :

\begin{Verbatim}[fontsize=\footnotesize, frame=single, framesep=1mm, framerule=0.1pt, rulecolor=\color{gray}]
mediatePI(Z, PI, Ls, (Z, PI, Avg)) :-
    findall(V, member((V,_),Ls), Values),
    avg(Values,AvgTmp),
    zone(Z, MediationPolicy), 
    propertyInstance(Z, PI, Prop, _, [Sensor]),
    sensorValue(Sensor, SensedValue),
    findValue(Policy, Prop, SensedValue, AvgTmp, Avg).
\end{Verbatim}
\vspace{-1.75mm}
\noindent
First {\tt\small mediatePI/4} averages all user requests for a specific {\tt\small propertyInstance} so to mediate possible user-user conflicts. Then, it exploit {\tt\small findValue/4} to mediate between the obtained average with the global policy enforce by the \admin. Such mediation is based on the {\tt\small MediationPolicy} of the wing (i.e. {\tt\small east}, {\tt\small west}), on the property type {\tt\small Prop} (i.e. {\tt\small light}, {\tt\small temp}) on the value obtained by the sensor of that instance (i.e. {\tt\small SensedValue}) and on the computed average value {\tt\small AvgTmp}.  

Fig.~\ref{fig:findValue} lists the code of predicate {\tt\small findValue/4}. 
The first clause of {\tt\small findValue/4} (lines 1--4), manages the temperature in both wings in the same way. After determining the current season (line 2), it enforces that the target value is within the season-dependent ranges specified for sustainability purposes (line 3--4), viz. 18--22$^\circ$C in Winter and Autumn and 28--24$^\circ$C in Summer and Spring. 
The second and the third clauses of {\tt\small findValue/4} (lines 5--6 and 7--10) manages instead the environmental brightness, depending on the wing of the room, on the current brightness and the weather. 
This process determines a mediated target state that reconciles all user-user and user-admin conflicts on each property instance, which fully answers \textbf{Q2}.

Finally, the answer to \textbf{Q3} is achieved through the implementation of the {\tt\small associateActions/2} predicate. 
In our \textit{Smart Building}, the policy we chose to adopt consists of dividing the workload equally between the various actuators, with the only exception of the heater that only accepts two values, viz. 0 or 100. The code is similar to the one proposed in the previous section for the \textit{Smart Home}, in which the {\tt\small selectActionsForPI/6} is adapted to the new policy and in case of multiple requests to the same actuator, now the maximum value is chosen.

\begin{Verbatim}[fontsize=\footnotesize, frame=single, framesep=1mm, framerule=0.1pt, rulecolor=\color{gray}]
selectActionsForPI(_, _, V, Actuators, _, Actions):-
    length(Actuators, L),triggerAll(V, L, Actuators, Actions).
    
triggerAll(_, _, [], []).
triggerAll(V, L, [A|Actuators], [(A,VNew)|Actions]):-
    dif(A, heater), 
    VNew is V/L, triggerAll(V, L, Actuators, Actions).
triggerAll(V, L, [heater|Actuators], [(heater,100)|Actions]):-
    V > 0, triggerAll(V, L, Actuators, Actions).
triggerAll(V, L, [heater|Actuators], [(heater,0)|Actions]):-
    V =< 0, triggerAll(V, L, Actuators, Actions).

setActuators(Actions, ExecutableActions) :- 
    setActuatorsWithMin(Actions, -inf,inf, ExecutableActions).
\end{Verbatim}
\vspace{-1.75mm}
\noindent
First {\tt\small selectActionsForPI/6} computes the number of actuators of the {\tt\small propertyInstance} (i.e. {\tt\small L}), then {\tt\small triggerAll/4} is called which distributes the workload to the actuators with the exception of the {\tt\small heater}. Finally, {\tt\small setActuators} choose the maximum in case of multiple sets for a specific actuator (with no lower or upper bound). With this process we can determine the correct configuration for each actuator acting on that state, and mediate between any conflicting configurations that a single actuator possibly receives, answering to \textbf{Q3}.

We conclude this section by describing a use case for the scenario above exploiting the policies described. Suppose that it is a sunny day in winter, with the brightness value sensed in the West common room at 160 out of 255, and user {\tt\small u1} sets the brightness of her room to 0 and the temperature to 18$^{\circ}$C. On the contrary, user {\tt\small 3} sets the temperature at 28$^{\circ}$. Assume that the two rooms share the A/C system but not the lighting system. Meanwhile, user {\tt\small u4} sets the temperature and brightness of {\tt\small room2}, which she is not authorised to handle. Finally, users {\tt\small u2} and {\tt\small u8} are both in the {\tt\small commonRoom\_E} with the same goal for the light but different goals for the temperature (respectively 23$^{\circ}$C and 18$^{\circ}$C).

\begin{Verbatim}[fontsize=\footnotesize, frame=single, framesep=1mm, framerule=0.1pt, rulecolor=\color{gray}]
season(winter).
sensorValue(lightCommonRoom_W, 160).
user(u1, [room_E_1, commonRoom_E, commonRoom_W]).
user(u2, [room_E_2, commonRoom_E, commonRoom_W]).
user(u3, [room_E_3, commonRoom_E, commonRoom_W]).
user(u4, [room_E_4, commonRoom_E, commonRoom_W]).
user(u8, [room_W_4, commonRoom_E, commonRoom_W]).
set(u1, room_E_1, roomLight, 0).
set(u1, room_E_1, roomTemp, 18).
set(u3, room_E_3, roomTemp, 28).
set(u4, room_E_2, roomLight, 0).
set(u4, room_E_2, roomTemp, 18).
set(u2, commonRoom_W, commonRoomLight, 255).
set(u2, commonRoom_W, commonRoomTemp, 23).
set(u8, commonRoom_W, commonRoomLight, 255).
set(u8, commonRoom_W, commonRoomTemp, 18).
\end{Verbatim}
\vspace{-1.75mm}
\noindent
For each room the result of the mediation phase with the application of global policies described before is:
\begin{Verbatim}[fontsize=\footnotesize, frame=single, framesep=1mm, framerule=0.1pt, rulecolor=\color{gray}]
[(room_E_1,roomLight,100), (room_E_1,roomTemp,18),
(room_E_3,roomTemp,22), (commonRoom_W,commonRoomLight,255),
(commonRoom_W,commonRoomTemp,20.5)]
\end{Verbatim}
\vspace{-1.75mm}
\noindent
where {\tt\small roomLight} of {\tt\small room\_E\_1} is bounded on 100 because it is the minimum bound for the light in the East wing, while the {\tt\small roomTemp} is within the bounds. Meanwhile, {\tt\small roomTemp} of {\tt\small room\_E\_3} is bounded to 22 the maximum temperature allowed in winter. Instead in {commonRoom\_W} the temperature is the average of the two requests because it is within the boundaries and also the light because we are in the West wing and it is sunny (brightness $>$ 100) so the maximum bound is 255, the goal of both users. Finally, the goals of user {\tt\small u4} are ignored because is not authorised to interact with {\tt\small room\_E\_2}. Then, the actions to be carried out, given the states computed, will be:
\begin{Verbatim}[fontsize=\footnotesize, frame=single, framesep=1mm, framerule=0.1pt, rulecolor=\color{gray}]
[(acCommonRoom_W, 20.5), (biglightCommonRoom_W_1, 127.5),  
(biglightCommonRoom_W_2, 127.5), (acOdd_E, 22), (heater, 100), 
(biglightRoom_E_1, 50), (smalllightRoom_E_1, 50)] 
 \end{Verbatim}
 \vspace{-1.75mm}
 \noindent
where the temperature of {\tt\small commonRoom\_W} is managed only by {\tt\small acCommonRoom\_W} and the light is implemented by two main light which equally divide the goal. Meanwhile, {\tt\small acOdd\_E} is the air conditioning system shared by room 1 and 3 and is setted to the maximum of the two goals (18 and 22) and also in {\tt\small room\_E\_1} the {\tt\small heater} is working. Finally, the small and big light work together to implement the goal.



\section{Related Work}
\label{sec:related}

\vspace{-1mm}
\noindent
In this section, we discuss some closely related work on the self-management of smart environments. Most of these works fall within three main categories, viz. \textit{goal-oriented}~\cite{goalorismartenv}, \textit{hierarchical}~\cite{hierarchicalgoal}, and \textit{neural and fuzzy}~\cite{neurofuzzyenergy}.

First, \cite{sembuild,autonomic,kratos,argumentation} and \cite{robots} propose goal-oriented approaches to conflict mediation.
%
Targeting global goals as energy efficiency, users comfort, and system security, \cite{sembuild} presents a solution to manage smart buildings by adding a semantic layer on top of the stack of IoT devices for reaching the desired global goals, exploiting an ontology of goal types.
With a more formal approach, \cite{autonomic} devise a methodology for autonomic device management describing the evolution of a smart environment as the set of evolutions of single device states, modelled as command sequences. Given a global goal, this solution determines the correct sequence of commands to reach it.
Besides, \cite{kratos} proposes an access control mechanism exploiting a priority-based policy negotiation technique to solve user-user conflicts in a smart home, made of multiple devices. 
%
%
Finally, Tartarus~\cite{robots}, is a Prolog platform designed to integrate cyber-physical systems and robots, supporting mobility, cloning, and payload carrying. 
More in general, \cite{argumentation} propose a solution for the problem of conflicts resolution in a multi-agent system, through argumentation-based reasoning. 
%
Naturally, as per its goal-oriented nature, \reasoner enables system administrators to write customised policies that can accommodate sophisticated and expressive mediation policies exploiting for example the semantic ontology described in~\cite{sembuild} or the negotiation technique proposed in~\cite{kratos}.

Second, hierarchical solutions for goal mediation have been proposed by~\cite{hierarchicalgoal} and~\cite{hierarchicalsmartoff}. Dynamic hierarchical goal management for different IoT systems is discussed in~\cite{hierarchicalgoal}, considering conflicting local and global goals, and the availability of limited resources that can vary at runtime. 
%
Regarding security in smart environments and in particular \textit{Smart Office}s, \cite{hierarchicalsmartoff} propose a hierarchical, agent-based solution that considers the high number of potential users, their security roles and the heterogeneity of devices and spaces. 
%
An interesting extension to \reasoner is to include hierarchical approaches to solve goals and to consider security aspects as well.

Third, and last, fuzzy logic~\cite{fuzzyhome,fuzzyexpert,fuzzycontroller} and neural network~\cite{smartenvnn,nnpreview2} approaches to goal mediation have been studied recently, along with their combination~\cite{neurofuzzypressure}.
Fuzzy logic can be used for context-awareness in Smart Home as illustrated in~\cite{fuzzyhome}, where raw data from the sensors are processed to manages actuators according to the computed context based on the user movement and activity.
The works in~\cite{fuzzyexpert,fuzzycontroller} propose expert systems to control the A/C of smart buildings, based on the current status of the sensors and the outside temperature. 
Similarly, \cite{neurofuzzypressure} manages A/C systems through a neuro-fuzzy controller where an adaptive neural network is used to better tuning the fuzzy rules, making them more robust. Neural networks have also been successfully used to predict energy consumption more reliably than traditional techniques~\cite{smartenvnn} and for indoor temperature forecasting~\cite{nnpreview2}. 
As \prolog is well-suited to implement fuzzy logic~\cite{rubio2014fuzzy}, an interesting extension of \reasoner is to accommodate fuzzy controllers. Similarly, predictions based on neural networks can be made available in the knowledge base of \reasoner from external services or by relying on recent implementations of \prolog that support neural networks (e.g. DeepProbLog~\cite{deepprobblog}).


\section{Concluding Remarks}
\label{sec:conclusions}

\noindent
This article proposed a declarative framework -- and its open-source Prolog prototype \reasoner~-- to specify policies for mediating contrasting (user and/or global) goals and actuator settings in smart environments. The prototype is provisioned as a service through \lpaas, and it can resolve user-user and user-admin conflicts into a target state for the smart-environment and a set of actuators settings to reach it. 

The wide variety of smart environments and the desiderata of their users and system administrators calls for new frameworks to easily develop and continuously adapt domain-specific mediation policies. 
This work moves some first steps towards this direction, aiming at contributing a novel declarative approach, enabled by \lpaas, to the field of goal-driven management of smart environments. 
As showcased in our example, thanks to its declarative nature, \reasoner features a suitable level of \textit{abstraction} and \textit{flexibility} to accommodate different needs of smart environments, making it easy to express, maintain and update mediation policies as per the ever-changing needs of IoT scenarios.
%

In our future work, we intend to:

\begin{itemize}

    \item \textit{New Policies. } Implement and test other mediation policies (e.g. based on fuzzy logic, learning or heuristics), by also proposing a set of \textit{building blocks} that \textit{System Administrators} can use to compose their own policies.
    \item \textit{Goal Geolocalisation. } Model a geo-localisation system for users and exploit machine learning to predict their movements and preferences, to reduce manual interaction.
    \item \textit{Web of Things. } Integrate \reasoner with Web of Things to make it more interoperable and easier to exploit in existing smart environments. 

\end{itemize}


\vspace{-2mm}
\bibliographystyle{ieeetr}
\bibliography{biblio}

\begin{thebibliography}{10}

\bibitem{iotfuture}
S.~K. Lee, M.~Bae, and H.~Kim, ``{Future of IoT Networks: A Survey},'' {\em
  Applied Sciences}, vol.~7, 2017.

\bibitem{iotreview}
Y.~Perwej, M.~A. AbouGhaly, B.~Kerim, and H.~A.~M. Harb, ``{An extended review
  on internet of things (iot) and its promising applications},'' {\em CAE},
  pp.~2394--4714, 2019.

\bibitem{iamhappy}
A.~Gyrard and A.~Sheth, ``{IAMHAPPY: Towards an IoT knowledge-based
  cross-domain well-being recommendation system for everyday happiness},'' {\em
  Smart Health}, vol.~15, p.~100083, 2020.

\bibitem{smartenvreview}
S.~{Merabti}, B.~{Draoui}, and F.~{Bounaama}, ``A review of control systems for
  energy and comfort management in buildings,'' in {\em ICMIC}, pp.~478--486,
  2016.

\bibitem{smarthomereview}
M.~R. {Alam}, M.~B.~I. {Reaz}, and M.~A.~M. {Ali}, ``{A Review of Smart Homes -
  Past, Present, and Future},'' {\em IEEE Trans. Syst. Man Cybern. Part C},
  vol.~42, no.~6, pp.~1190--1203, 2012.

\bibitem{smartenvreviewenergy}
E.~Torunski, R.~Othman, M.~Orozco, and A.~E. Saddik, ``A review of smart
  environments for energy savings,'' {\em ANT/MobiWIS}, vol.~10, pp.~205 --
  214, 2012.

\bibitem{reasoningconflicts}
H.~Sfar, B.~Raddaoui, and A.~Bouzeghoub, ``{Reasoning Under Conflicts in Smart
  Environment},'' in {\em ICONIP (3)}, pp.~924--934, 2017.

\bibitem{rulebased}
T.~{Perumal}, M.~N. {Sulaiman}, S.~K. {Datta}, T.~{Ramachandran}, and C.~Y.
  {Leong}, ``{Rule-based conflict resolution framework for Internet of Things
  device management in smart home environment},'' in {\em GCC£}, pp.~1--2,
  2016.

\bibitem{fuzzyexpert}
A.~Salih, ``{Fuzzy Expert Systems to Control the Heating, Ventilating and Air
  Conditioning (HVAC) Systems},'' {\em IJERT}, vol.~4, 2015.

\bibitem{masapproach}
P.~Davidsson and M.~Boman, ``{Saving Energy and Providing Value Added Services
  in Intelligent Buildings: A MAS Approach},'' in {\em ASA/MA}, pp.~166--177,
  2000.

\bibitem{semioticmas}
D.~Booy, K.~Liu, B.~Qiao, and C.~Guy, ``{A Semiotic Multi-Agent System for
  Intelligent Building Control},'' {\em AMBI-SYS}, 2008.

\bibitem{smartenvnn}
R.~Kumar, R.~Aggarwal, and J.~Sharma, ``{Energy analysis of a building using
  artificial neural network: A review},'' {\em Energy \& Buildings},
  pp.~352--358, 2013.

\bibitem{ifttt}
IFTT, ``{IFTTT: If This Then That}.'' https://ifttt.com/.

\bibitem{alexa}
Amazon, ``{What Is Alexa?}.'' https://developer.amazon.com/en-US/alexa.

\bibitem{surveyzambonelli}
C.~Becker, C.~Julien, P.~Lalanda, and F.~Zambonelli, ``{Pervasive computing
  middleware: current trends and emerging challenges},'' {\em CCF Trans.
  Pervasive Comput. Interact. 1}, vol.~1, pp.~10--23, 2019.

\bibitem{lpaas}
R.~{Calegari}, E.~{Denti}, S.~{Mariani}, and A.~{Omicini}, ``{Logic programming
  as a service},'' {\em {Theory and Pract. Log. Program.}}, vol.~18, no.~5-6,
  p.~846–873, 2018.

\bibitem{lpaasiot}
R.~{Calegari}, E.~{Denti}, S.~{Mariani}, and A.~{Omicini}, ``{Logic Programming
  as a Service (LPaaS): Intelligence for the IoT},'' in {\em ICNSC},
  pp.~72--77, 2017.

\bibitem{goalorismartenv}
J.~Palanca, E.~Del~Val, A.~Garcia-Fornes, H.~Billhardt, J.~M. Corchado, and
  V.~Juli{\'a}n, ``{Designing a goal-oriented smart-home environment},'' {\em
  Inf. Syst. Frontiers}, vol.~20, no.~1, pp.~125--142, 2018.

\bibitem{hierarchicalgoal}
A.~{Jantsch et al.}, ``{Hierarchical dynamic goal management for IoT
  systems},'' in {\em ISQED}, pp.~370--375, 2018.

\bibitem{neurofuzzyenergy}
S.~Naji, S.~Shamshirband, H.~Basser, A.~Keivani, U.~J. Alengaram, M.~Z. Jumaat,
  and D.~Petkovi{\'c}, ``{Application of adaptive neuro-fuzzy methodology for
  estimating building energy consumption},'' {\em Renew. and Sustain. En.
  Reviews}, vol.~53, pp.~1520--1528, 2016.

\bibitem{sembuild}
A.~{Andrushevich}, M.~{Staub}, R.~{Kistler}, and A.~{Klapproth}, ``{Towards
  semantic buildings: Goal-driven approach for building automation service
  allocation and control},'' in {\em ETFA 2010}, pp.~1--6, 2010.

\bibitem{autonomic}
M.~Sanaullah, F.~Corno, and F.~Razzak, ``{Autonomic goal-oriented device
  management for Smart Environments},'' {\em J. Ambient Intell. Smart
  Environ.}, vol.~7, no.~4, pp.~425--448, 2015.

\bibitem{kratos}
A.~K. Sikder, L.~Babun, Z.~B. Celik, A.~Acar, H.~Aksu, P.~McDaniel, E.~Kirda,
  and A.~S. Uluagac, ``{Kratos: Multi-User Multi-Device-Aware Access Control
  System for the Smart Home},'' WiSec '20, p.~1–12, 2020.

\bibitem{argumentation}
Z.~Shams, M.~D. Vos, N.~Oren, and J.~A. Padget, ``{Argumentation-Based
  Reasoning about Plans, Maintenance Goals, and Norms},'' {\em {ACM} Trans.
  Auton. Adapt. Syst.}, vol.~14, no.~3, pp.~9:1--9:39, 2020.

\bibitem{robots}
T.~Semwal, M.~Bode, V.~Singh, S.~S. Jha, and S.~B. Nair, ``{Tartarus: a
  multi-agent platform for integrating cyber-physical systems and robots},'' in
  {\em {AIR}}, pp.~20:1--20:6, {ACM}, 2015.

\bibitem{hierarchicalsmartoff}
I.~Mars{\'a}-Maestre, E.~De~La~Hoz, B.~Alarcos, and J.~R. Velasco, ``{A
  Hierarchical, Agent-based Approach to Security in Smart Offices},'' in {\em
  ICUC}, 2006.

\bibitem{fuzzyhome}
A.~{Patel} and T.~A. {Champaneria}, ``{Fuzzy logic based algorithm for Context
  Awareness in IoT for Smart home environment},'' in {\em TENCON},
  pp.~1057--1060, 2016.

\bibitem{fuzzycontroller}
F.~Calvino, M.~{La Gennusa}, G.~Rizzo, and G.~Scaccianoce, ``The control of
  indoor thermal comfort conditions: introducing a fuzzy adaptive controller,''
  {\em Energy \& Buildings}, vol.~36, no.~2, pp.~97 -- 102, 2004.

\bibitem{nnpreview2}
N.~Attoue, I.~Shahrour, and R.~Younes, ``{Smart building: Use of the artificial
  neural network approach for indoor temperature forecasting},'' {\em
  Energies}, vol.~11, no.~2, p.~395, 2018.

\bibitem{neurofuzzypressure}
W.~Jian and C.~Wenjian, ``{Development of an adaptive neuro-fuzzy method for
  supply air pressure control in HVAC system},'' in {\em SMC}, pp.~3806--3809,
  2000.

\bibitem{rubio2014fuzzy}
C.~Rubio{-}Manzano and P.~J. Iranzo, ``{A Fuzzy Linguistic Prolog and its
  Applications},'' {\em J. Intell. Fuzzy Syst.}, vol.~26(3), no.~3,
  pp.~1503--1516, 2014.

\bibitem{deepprobblog}
R.~Manhaeve, S.~Dumancic, A.~Kimmig, T.~Demeester, and L.~D. Raedt,
  ``{DeepProbLog: Neural Probabilistic Logic Programming},'' {\em CoRR},
  vol.~abs/1907.08194, 2019.

\end{thebibliography}

\end{document}